# Epitaxial Electrodeposition of Fe with Controlled In-Plane Variants for Reversible Metal Anode in Aqueous Electrolyte

Chenxi Sui[1], Ching-Tai Fu[2], Guangxia Feng[1], Yuqi Li[1], Junyan Li[1], Gangbin Yan[3], Po-Chun Hsu[2], Steven Chu[3,4,5], Yi Cui[1,4,6]*

[1]Department of Materials Science and Engineering, Stanford University, Stanford, CA

[2]Pritzker School of Molecular Engineering, University of Chicago, Chicago, IL, USA

[3]Department of Physics, Stanford University, Stanford, CA, USA

[4]Department of Energy Science and Engineering, Stanford University, Stanford, CA, USA

[5]Department of Molecular and Cellular Physiology, Stanford University, Stanford, CA, USA

[6]Stanford Institute for Materials and Energy Sciences, SLAC National Accelerator Laboratory, 2575 Sand Hill Road, Menlo Park, CA, USA.

*Corresponding Author: Yi Cui (yicui@stanford.edu)

**Abstract**

The development of reversible metal anodes is a key challenge for advancing aqueous battery technologies, particularly for scalable and safe stationary energy storage applications. Here we demonstrate a strategy to realize epitaxial electrodeposition of iron (Fe) on single-crystal copper (Cu) substrates in aqueous electrolytes. We compare the electrodeposition behavior of Fe on polycrystalline and single-crystalline Cu substrates, revealing that the latter enables highly uniform, dense, and crystallographically aligned Fe growth. Comprehensive electron backscatter diffraction (EBSD) and X-ray diffraction (XRD) analysis confirms the formation of Fe with specific out-of-plane and in-plane orientations, including well-defined rotational variants. Our findings highlight that epitaxial electrodeposition of Fe can suppress dendritic growth and significantly enhance Coulombic efficiency during plating/stripping cycles. This approach bridges fundamental crystallography with practical electrochemical performance, providing a pathway toward high-efficiency aqueous batteries utilizing Earth-abundant materials.

**Introduction**

Aqueous batteries are gaining attention for scalable stationary energy storage due to their intrinsic safety, low cost, and environmental compatibility. (*1-7*) Among various chemistries, iron-based aqueous batteries are especially attractive given the abundance and affordability of iron, (*8*) along with the well-established infrastructure of the steel industry that enables scalable implementation. Despite these advantages, iron metal anodes face critical challenges in achieving reversible electrochemical cycling. (*8-13*) Issues such as dendritic growth, hydrogen evolution, and low Coulombic efficiency hinder their technical advancement. Improving the morphology and reversibility of iron plating/stripping is essential for enabling high-performance aqueous iron batteries.

Epitaxial electrodeposition offers a potential solution by promoting crystallographically aligned growth of metal layers. (*8, 13-21*) By matching the deposited metal's orientation with that of the substrate, epitaxy can reduce nucleation barriers, suppress dendrites, and enable more uniform deposition. Electrochemical epitaxy differs fundamentally from traditional vapor-phase epitaxy in both mechanism and implications. In vapor-phase epitaxy, atoms arrive on the substrate in a high-energy, uncoordinated state and require high-temperature annealing to diffuse into thermodynamically favorable lattice positions. This process is typically irreversible and sensitive to kinetic trapping. In contrast, electrochemical epitaxy involves redox-active species such as $Fe^{2+}$ ions that are surrounded by hydration shells in solution. These hydrated ions may already align with crystallographically favorable sites at the moment of reduction, facilitating low-barrier nucleation and oriented growth. Moreover, the electrochemical process is inherently reversible, allowing for controlled cycling of Fe deposition and dissolution. This distinction highlights electrochemical epitaxy as a unique and underutilized pathway to achieving highly ordered, functional interfaces under ambient conditions. (Table 1)

In this work, we demonstrate that aqueous electrodeposition of Fe on single-crystalline Cu substrates can result in epitaxial growth with well-defined crystallographic orientation and in-plane alignment. Through comprehensive structural characterization, we reveal the existence of specific orientation relationships and twinning phenomena. Electrochemical tests show that such epitaxial control enhances plating/stripping reversibility and suppresses dendrite formation. Our study presents a platform for integrating crystallographic control into aqueous battery design, bridging the gap between materials science and practical energy storage technology.

| Epitaxy type | Traditional (vapor phase) epitaxy | Electrochemical epitaxy |
|---|---|---|
| Representation<br><br>⚪ Fe<br>🟢 Transition state Fe<br>🔴 Undiffused Fe | 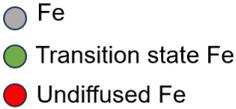 | 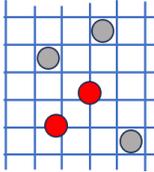 |
| Epitaxy mechanism | Need annealing to allow Fe diffuse to thermodynamically favorable sites | $Fe^{2+}$ surrounded by hydration shells, may have already diffused to the corresponding sites |
| Redox chemical reaction | no | yes, $Fe^{2+}$ to Fe |
| Hydration shell | no | yes |
| Reversibility | Not reversible | reversible |

Table 1. Comparison between traditional vapor phase epitaxy and electrochemical epitaxy.

## Mechanism of Electrochemical Epitaxy

Variations in atomic arrangement and crystallographic orientation across substrates significantly influence the morphology and crystallinity of electrodeposited metals. On polycrystalline substrates, where grain orientations are randomly distributed and surface atomic configurations are heterogeneous, metal nucleation tends to occur stochastically. This often results in irregular growth, poor crystallinity, and the formation of dendritic structures. Dendrites increase the electrode's surface area, which enhances undesired side reactions such as hydrogen evolution. Additionally, dendritic deposits can mechanically destabilize the electrode by penetrating the separator or detaching during stripping, thereby reducing the electrochemical reversibility of the system.

In contrast, when the substrate is engineered to be textured or single-crystalline, with uniform lattice orientation and well-ordered surface atoms, electrodeposition proceeds in a more controlled and uniform manner. Under such conditions, the deposited metal can adopt a preferential orientation, and in favorable cases, epitaxial growth is achieved (Figure 1a). This structural coherence at the interface contributes to smoother morphology, reduced dendrite formation, and improved cycling stability.

To highlight this effect, we performed electrodeposition of Fe on both polycrystalline Cu and single-crystal Cu (100) substrates. The XRD 2θ scan of Fe deposited on polycrystalline Cu (Figure 1b) reveals multiple diffraction peaks—Fe (110), Fe (211), and Fe (222)—indicative of randomly oriented grains. This structural disorder is reflected in the rough and inhomogeneous surface morphology observed in SEM images (Figure 1d). In contrast, Fe electrodeposited on Cu (100) displays a single Fe (110) diffraction peak (Figure 1c), consistent with a highly oriented film. The corresponding SEM image (Figure 1e) shows a dense and uniform morphology, characteristic of coherent growth guided by the underlying substrate.

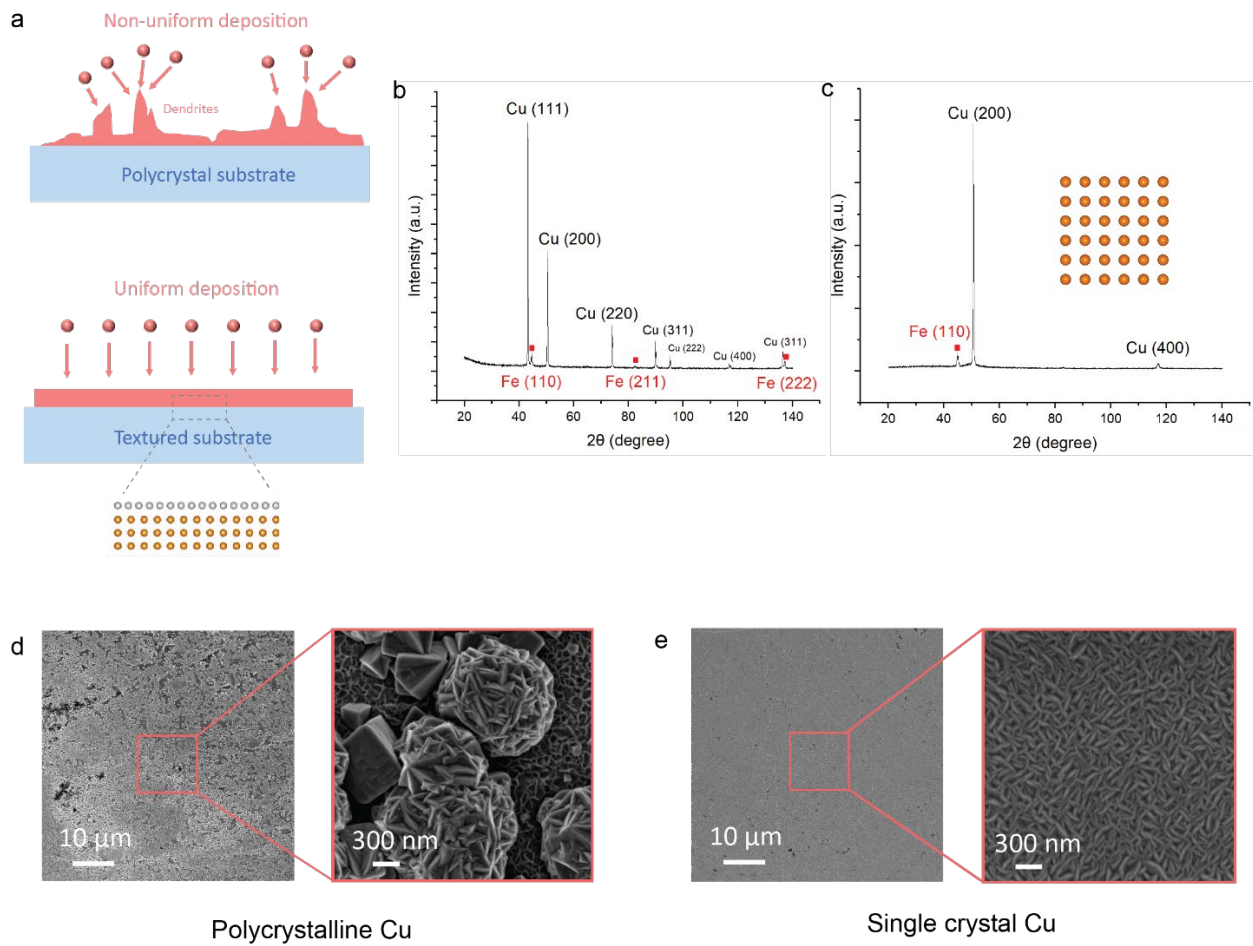

Figure 1. Electrodeposition of Fe on polycrystalline and single-crystal Cu substrates. (a) Schematic illustration comparing metal deposition morphology on polycrystalline versus textured (e.g., single-crystalline) Cu substrates. (b, c) XRD 2θ scans of Fe electrodeposited on polycrystalline Cu and single-crystal Cu (100), respectively. The inset in (c) shows the atomic arrangement of the Cu (100) surface. (d, e) SEM images of Fe films deposited on polycrystalline and single-crystal Cu substrates, respectively, highlighting the improved uniformity and compactness achieved with epitaxial growth.

**XRD Characterization of Epitaxial Fe**

Inspired by the morphology uniformity improvement of using single crystal Cu as the substrate, we went further to study the Fe electrodeposition on single crystal Cu substrates with three different orientations of (100), (110), (111), respectively. To evaluate the crystallographic quality of electrodeposited Fe films at a larger spatial scale, X-ray diffraction (XRD) was employed. Compared to EBSD, which typically probes micron-scale regions, the XRD beam spot size (~6 mm) enables more global structural characterization. Figure 2a presents the 2θ scans of Fe films deposited on single-crystal Cu substrates with different orientations. For both Cu (111) and Cu (100) substrates, the Fe films exhibit a dominant (110) out-of-plane orientation, indicating preferential crystallographic alignment. In contrast, Fe electrodeposited on Cu (110) shows a dominant Fe (211) peak, suggesting that the substrate orientation plays a critical role in guiding film texture. To illustrate ideal epitaxial single-crystal growth, atomic schematics of Fe epitaxy on Cu are shown next to each diffraction pattern. More detailed atomic arrangements can be found in Supplementary Figs. 3, 4, and 5. To further assess the full crystallographic orientation, particularly the in-plane alignment, we performed XRD pole figure measurements for all three substrate types. These were compared with simulated pole figures of ideal single-crystalline Fe (110) and Fe (211) (Figures 2e and 2f) to identify the degree of twinning and the presence of rotational variants. On Cu (111), the Fe (110) pole figure (Figure 2b) reveals three sets of well-defined poles arranged with 60° in-plane rotation, consistent with the sixfold symmetry and the formation of three rotational variants. On Cu (100), Fe (110) is again the preferred orientation, but the pole figure (Figure 2d) shows two rotational variants separated by 90°, reflecting the underlying fourfold symmetry of the Fe. Remarkably, electrodeposition on Cu (110) yields Fe films with a high-index (211) orientation. The corresponding pole figure (Figure 2c) exhibits two sets of Fe (211) poles with 180° rotational symmetry, indicating the presence of twin domains with well-defined orientation. Although the Fe deposits on the three Cu substrates are not perfect single crystals, they exhibit well-defined twinning and in-plane rotational variants, reflecting a high degree of crystallographic order and predictable epitaxial behavior.

These observations underscore the capacity of electrochemical epitaxy to produce highly regulated, crystallographically aligned Fe films, including those with high-index orientations that are typically difficult to synthesize. The presence of controlled twinning and discrete in-plane variants further highlights the epitaxial nature of the growth, even in a body-centered cubic (BCC) system where such order is rarely observed under ambient conditions.

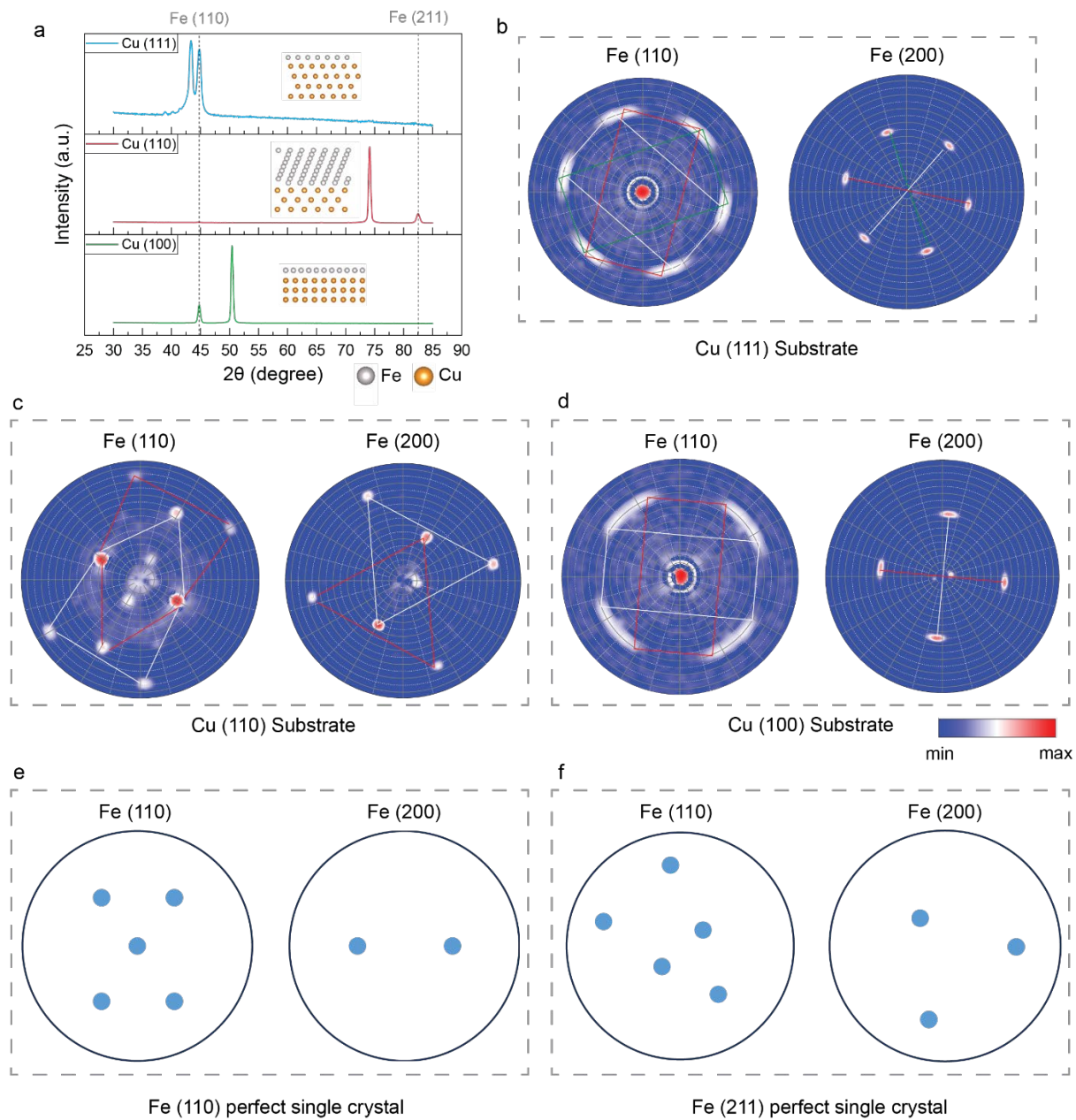

Figure 2. Crystallographic characterization of electrodeposited Fe on different single-crystal Cu substrates. (a) XRD 2θ scans of Fe films electrodeposited on Cu (111), Cu (100), and Cu (110), showing the preferred out-of-plane orientations. The three insets adjacent to the diffraction patterns illustrate the atomic arrangements of Fe (110) on Cu (111), Fe (211) on Cu (110), and Fe (110) on Cu (100), respectively (b–d) Experimental XRD pole figures of Fe deposited on Cu (111), Cu (110), and Cu (100), respectively. Each sample includes two pole figures corresponding to the Fe (110) and Fe (200) 2θ angles. Colored lines are used to highlight individual sets of epitaxial orientations; each color denotes one distinct rotational variant of single-crystal Fe. (e, f) Simulated pole figures of ideal single-crystal Fe with (110) and (211)

orientations, used as references for identifying twinning and in-plane variants in the experimental data.

**Morphology and Crystallinity of the Electrodeposition**

Figure 3a confirms the strong out-of-plane crystallographic alignment of the annealed single-crystal Cu substrate. Electrochemical measurements (Figure 3b) reveal that Fe deposited on polycrystalline Cu exhibits a significantly higher overpotential compared to Fe deposited on single-crystal Cu substrates, indicating that a well-textured substrate facilitates more favorable Fe nucleation and growth. SEM images (Figures 3c, 3f, and 3i) show that Fe electrodeposited on all single-crystal Cu orientations forms dense, uniform, and well-regulated morphologies. Crystallographic orientation mapping using EBSD inverse pole figures (Figures 3d, 3g, and 3j) indicates that Fe deposited on Cu (111) and Cu (100) adopts a dominant (110) out-of-plane orientation, while deposition on Cu (110) results in a (211) orientation. To assess in-plane orientation, EBSD pole figures were generated (Figures 3e, 3h, and 3k), showing excellent agreement with the corresponding XRD pole figures in Figure 2. Notably, the EBSD pole figures exhibit higher clarity, likely due to the smaller probe size of EBSD, which minimizes the averaging over defects and misorientations compared to XRD. The symmetry of the in-plane patterns also reflects the morphology growth direction. Fe grown on Cu (111) exhibits threefold symmetry, consistent with three equivalent in-plane growth directions separated by 120°, as observed in Figure 3c. On Cu (100), Fe grows along two orthogonal directions separated by 90°, corresponding to the substrate's fourfold symmetry (Figure 3i). Inverse pole figures and pole figures for all Fe orientations are shown in Supplementary Fig. 1. These growth directions are highlighted with arrows in the respective SEM images. This integrated structural and morphological analysis reveals a strong correlation between substrate crystallographic orientation and the resulting Fe deposition morphology.

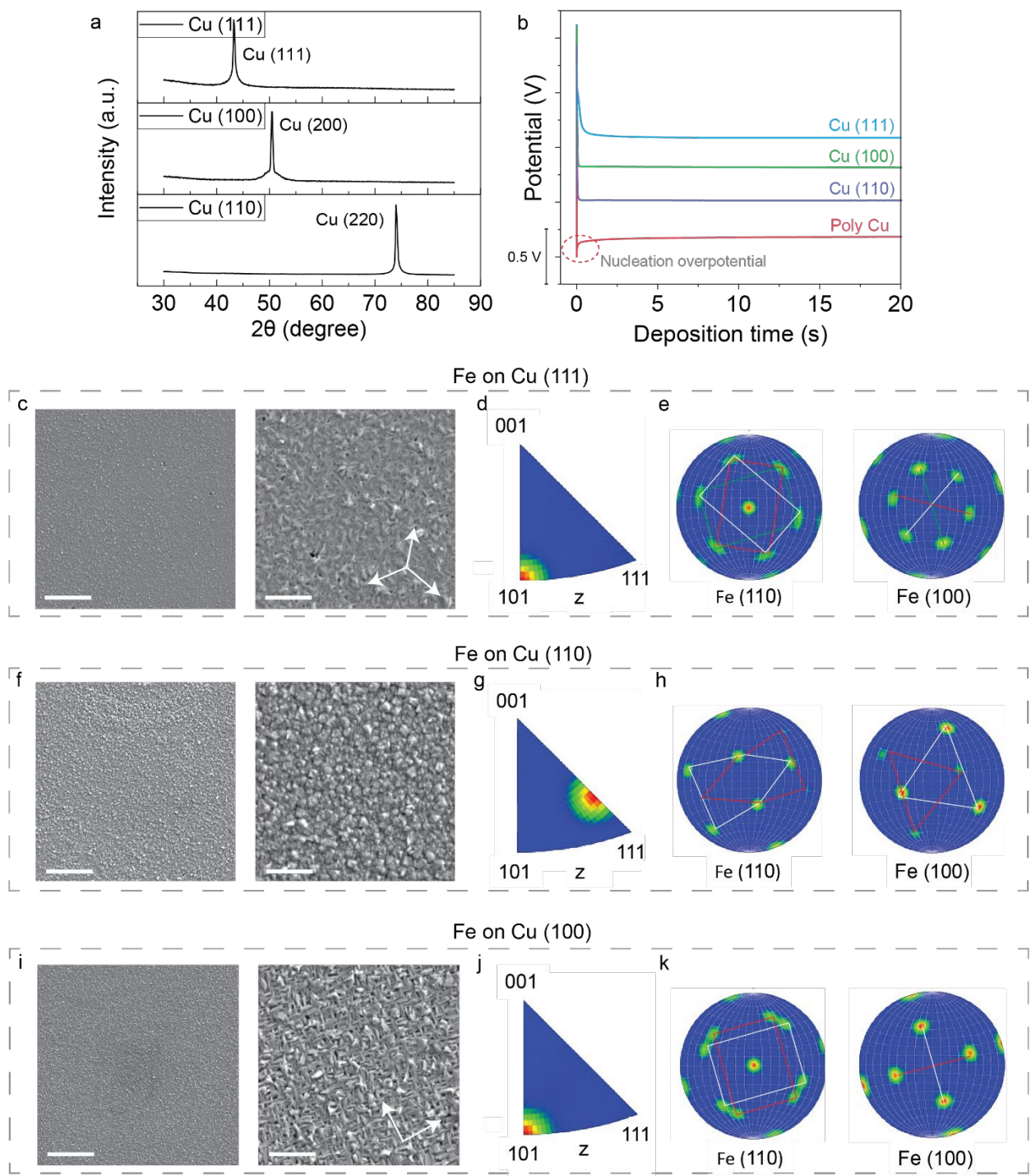

Figure 3. Characterization of Fe electrodeposition. (a) XRD 2θ scan of single-crystal Cu. (b) Voltage-time curves of Fe electrodeposition under constant current, 10 mA cm$^{-2}$. (c, f, i) SEM photos of electrodeposited Fe on single-crystal Cu (111), Cu (110), and Cu (100). Scale bars in right photos are 5 μm and 500 nm. (d, g, j) EBSD inverse pole figures and (e, h, k) EBSD pole figures of electrodeposited Fe on single crystal Cu (111), Cu (110), and Cu (100).

**Reversibility of electro-plating/stripping in aqueous electrolyte**

Following the demonstration of high-quality Fe electrodeposition on single-crystal Cu substrates, we explored a more cost-effective approach to obtain crystallographically aligned Cu by annealing commercial polycrystalline Cu foils (Figure 4a). The annealing method follows our previous work. (*17*) Although the annealed Cu is not a perfect single crystal—as evidenced by the presence of satellite intensities around the main poles in the XRD pole figure (Figure 4b), which indicates the existence of crystallographic defects and minor misorientations—it still exhibits significant texturing. This suggests that while full single crystallinity is not achieved, the substrate retains sufficient orientational coherence to influence subsequent Fe deposition. From a practical perspective, perfect single crystallinity is not a prerequisite for high-performance battery electrodes. Given the trade-off between fabrication cost and electrochemical performance, a well-textured Cu substrate is often sufficient to achieve substantially improved reversibility in metal plating/stripping processes.

To assess the electrochemical impact of substrate crystallinity, we assembled Fe‖Cu asymmetric coin cells using both polycrystalline and annealed single-crystal Cu electrodes (Figure 4c). Galvanostatic cycling at a current density of 0.5 mA cm$^{-2}$ reveals that the annealed single-crystal Cu electrode delivers a higher Coulombic efficiency (~96%) compared to its polycrystalline counterpart (~91%). Detailed charge–discharge profiles (Figures 4e and 4f) show that the polycrystalline Cu exhibits a larger overpotential during stripping, while the single-crystal Cu electrode maintains more stable voltage behavior, consistent with more uniform Fe deposition and dissolution. Finally, long-term cycling was performed using the Cu (100) single-crystal substrate to evaluate durability (Figure 4g). The coin cell assembled with a polycrystalline Cu electrode short-circuited after approximately 100 cycles due to non-uniform deposition and dendrite-induced failure. In contrast, the cell with the Cu (100) single-crystal electrode maintained stable performance for over 200 cycles, with no observable degradation and consistently high Coulombic efficiency around 96%.

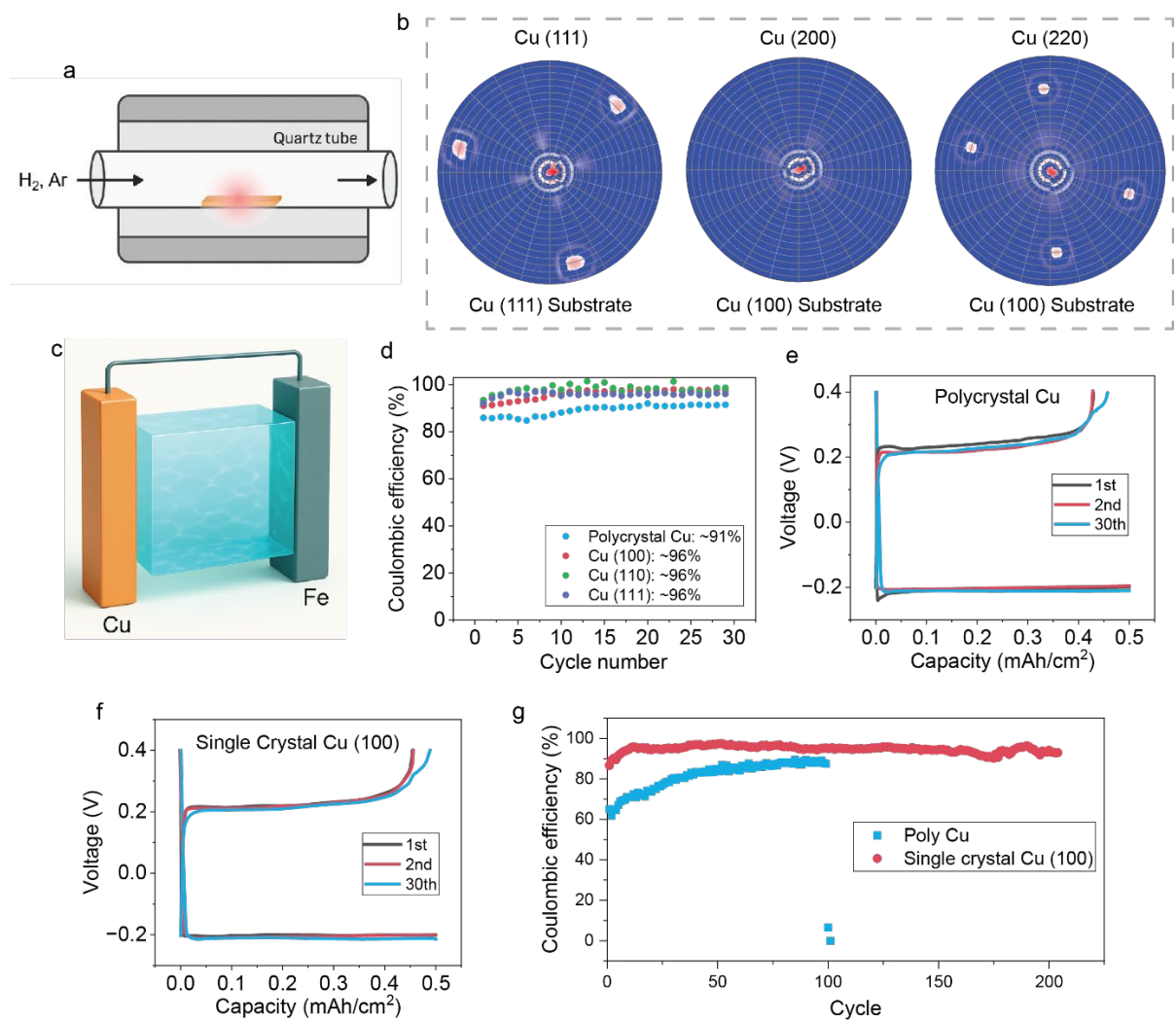

Figure 4. Electrochemical reversibility of Fe || Cu asymmetric coin cells using different Cu substrates. (a) Schematic illustration of the annealing process used to convert polycrystalline Cu into textured Cu. (b) XRD pole figures of annealed Cu foil, indicating the formation of a highly textured (though not perfect single-crystal) surface. (c) Schematic of the Fe || Cu coin cell configuration used for reversibility testing. (d) Coulombic efficiency comparison of Fe || Cu cells assembled with polycrystalline Cu and single-crystal Cu(100) electrodes, cycled at 0.5 mA cm$^{-2}$. The cut-off voltage is set at 0.4 V. (e, f) Representative charge-discharge voltage profiles for cells with polycrystalline Cu and single-crystal Cu(100) electrodes, respectively. (g) Long-term Coulombic efficiency stability for Fe || Cu cells showing superior performance and cycling durability with the single-crystal Cu(100) electrode. The cut-off voltage is set at 0.5 V.

## Discussion

This work demonstrates that epitaxial electrodeposition of Fe on single-crystal and textured Cu substrates provides a promising pathway to achieving structurally uniform, highly reversible metal anodes in aqueous electrochemical systems. Through detailed crystallographic and morphological analyses, we show that single-crystal Cu facilitates the formation of Fe films with well-defined (110) and (211) orientations and in-plane rotational variants, leading to smoother deposition, suppressed dendrite growth, and improved Coulombic efficiency compared to conventional polycrystalline substrates.

Although this study focused primarily on the materials science aspects—elucidating the growth behavior, orientation control, and structural quality of electrochemically deposited Fe—our findings highlight the broader potential of electrochemical epitaxy in aqueous battery applications. The strategy could be particularly valuable for future anode-free battery designs, where textured metal foils serve as current collectors to guide uniform, reversible metal plating during cycling.

Looking forward, a critical challenge lies in scaling up the production of low-cost, large-area textured metal substrates. While single-crystal Cu offers an ideal platform for fundamental investigation, practical implementation will require accessible methods such as thermal annealing or directional solidification to produce industrially viable textured foils. Advancing these capabilities is essential for translating the concept of electrochemical epitaxy into sustainable, scalable technologies for next-generation aqueous batteries.

## Method

### Materials

The single crystal Cu substrates are purchased from MSE Supplies Inc. $FeSO_4$ and $(NH_4)_2SO_4$ are purchased from Sigma Aldrich. The aqueous electrolyte is synthesized by mixing 2M $FeSO_4$, and 0.1M $(NH_4)_2SO_4$ in DI water. The $(NH_4)_2SO_4$ is used to improve the uniformity of the electrodeposition. (Supplementary Fig. 2) The electrodeposition to grow Fe is done in the water glove box.

### Characterization

The SEM and EBSD characterizations were conducted by a TESCAN LYRA3 field-emission scanning electron microscope (FESEM) with an Oxford Instruments NordlysMax2 electron backscatter diffraction detector at The University of Chicago. The XRD and pole figure were acquired by PANalytical X'Pert 2 in Stanford Nano Shared Facilities. The theoretical pole figures were calculated by an open-source software *ReciPro* (https://github.com/seto77/ReciPro).

### Textured Cu annealing

Single-crystal Cu substrates with (111), (100), and (110) orientations were synthesized via hydrogen-assisted annealing of commercial Cu foils or disks in a quartz tube furnace. For Cu (111), a 25 μm-thick Cu foil was annealed at 1030 °C for 10 hours under a 1:1 $H_2$/Ar mixture (70 sccm each), followed by natural cooling. Cu (100) was prepared by annealing a 3 cm², 1 mm-thick Cu disk at 1070 °C for 10 hours under a 1:9 $H_2$/Ar mixture (100/900 sccm), followed by slow cooling to 900 °C at 1 °C min$^{-1}$ and natural cooling. Cu (110) was synthesized by annealing a similar Cu disk at 1030 °C for 10 hours under 30 sccm $H_2$ and 70 sccm Ar, with vacuum pumping applied throughout the process.

**Electrochemistry**

The electrodeposition is conducted by a VMP3 potentiostat (BioLogic). The current density is set as 10 mA cm$^{-2}$. The half-cell performance was tested in a coin cell (CR2032) using glassy fiber as a separator and was run by the Land BT2000 battery test system. The cell was cycled under 0.5 mA cm$^{-2}$.

**Contributions**

Y.C. and C.S. conceived the idea. C.S. performed all the experiments and the corresponding data analyses. C.-T.F. and C.S. did the EBSD. C.S. and Y.C. wrote the manuscript with input from all co-authors.

**Acknowledgments**

This work is supported by the Aqueous Battery Consortium, an energy innovation hub under the U.S. Department of Energy, Office of Basic Energy Sciences, Division of Materials Science and Engineering. Part of this work was performed at the Stanford Nano Shared Facilities (SNSF) and Stanford Nanofabrication Facility (SNF).

**Data availability**

All data needed to support the conclusions in the paper are present in the manuscript and/or Supplementary Information. Additional data related to this paper may be requested from the corresponding authors.